\documentclass[journal,twoside,web]{ieeecolor}
\usepackage{lcsys}
\usepackage{cite}
\usepackage{amsmath,amssymb,amsfonts}
\usepackage{algorithmic}
\usepackage{graphicx}
\usepackage{textcomp}
\usepackage{amsmath}
\usepackage{color}

\usepackage{amssymb}  % assumes amsmath package installed. This line's from the prev, ieeeconf version.

\usepackage{mathtools,subcaption,float} % I added these, they're not mentioned in the docs - Raph

\newcommand{\afn}{{g_f^-}}
\newcommand{\asp}{{g_s^+}}
\newcommand{\asn}{{g_s^-}}
\newcommand{\ausp}{{g_{us}^+}}

\newcommand{\torque}{I}

\newcommand{\iappi}{{i_{\rm{app},i}}}
\newcommand{\iappnominal}{{i_{\rm{nominal}}}}

\newcommand{\wref}{{\omega_{\rm{ref}}}}

\newcommand{\Aref}{{A_{\rm{ref}}}}

\newcommand{\pw}{p_{\omega}}
\newcommand{\pA}{p_{A}}

\newcommand{\qsensory}{q_p}
	
\begin{document}
\title{\LARGE \bf
	Neuromorphic Control of a Pendulum
}

\author{Raphael Schmetterling, Fulvio Forni, Alessio Franci and Rodolphe Sepulchre
	\thanks{The research leading to these results has received funding from the European Research Council under the
		Advanced ERC Grant Agreement SpikyControl n.101054323. R. Schmetterling has received funding from an EPSRC Doctoral Training Programme [grant number EP/R513180/1].}% <-this % stops a space
	\thanks{R. Schmetterling, F. Forni and R. Sepulchre are with the Department of Engineering, University of Cambridge, Trumpington Street, Cambridge CB2 1PZ, United Kingdom (e-mails: rjzs2@cam.ac.uk, f.forni@eng.cam.ac.uk and r.sepulchre@eng.cam.ac.uk).}
	\thanks{A. Franci is with the Department of Electrical Engineering and Computer Science, University of Liège, and WEL Research Institute, Wavre, Belgium (e-mail: afranci@uliege.be).}% 
	\thanks{R. Sepulchre is also with the Department of Electrical Engineering, KU Leuven, KasteelPark Arenberg, 10, B-3001 Leuven, Belgium (e-mail: rodolphe.sepulchre@kuleuven.be).}
	%Dayton, OH 45435, USA
	%{\tt\small b.d.researcher@ieee.org}}%
	}

\maketitle
\thispagestyle{empty}  % Removes the page number in the first page
\pagestyle{empty}

%%%%%%%%%%%%%%%%%%%%%%%%%%%%%%%%%%%%%%%%%%%%%%%%%%%%%%%%%%%%%%%%%%%%%%%%%%%%%%%%
\begin{abstract}
	
	We illustrate the potential of neuromorphic control on the simple mechanical model of a pendulum, with both event-based actuation and sensing. The controller and the pendulum are regarded as event-based systems that occasionally interact to coordinate their respective rhythms. Control occurs through a proper timing of the interacting events. We illustrate the mixed nature of the control design: the design of a rhythmic automaton, able to generate the right sequence of events, and the design of a feedback regulator, able to tune the timing of events. 
	
\end{abstract}

%%%%%%%%%%%%%%%%%%%%%%%%%%%%%%%%%%%%%%%%%%%%%%%%%%%%%%%%%%%%%%%%%%%%%%%%%%%%%%%%

\section{Introduction}

% IEEE advice on graphics:
% https://journals.ieeeauthorcenter.ieee.org/create-your-ieee-journal-article/create-graphics-for-your-article/

%% RESUBMISSION DEADLINE: 12th May.

\IEEEPARstart{M}{ost} robotic tasks can be decomposed as sequences of events \cite{lakatos2014nonlinear,williamson1999robot}. 
This is best illustrated in the context of animal-like movements such as walking \cite{garofalo2012walking} or swimming \cite{ijspeert2007swimming}. In spite of spectacular advances in robotics, animals still drastically overperform robots in  performing such tasks reliably in changing and uncertain environments. The vision of neuromorphic engineering is that the event-based nature of animal control is key to its superior performance over our current clocked digital technology \cite{mead1989analog}.  
%Event-based technology has developed at a fast pace in the recent decade: event-based cameras are opening a new age in computer vision \cite{gallego2020event,gothard2022digital}, while spiking neural networks offer new perspective in remedying the current energy inefficiency of AI \cite{tavanaei2019deep}. 

The potential of event-based control was recognized from the early days of neuromorphic engineering \cite{deweerth1991simple}. 
%Åström and colleagues empirically explored the potential of a spiking controller in a simple DC motor servo system, illustrating the ability of the controller to interpolate between the behavior of a Pulse-Width-Modulated controller at nominal speed and the behavior of a stepper motor at very low speed. 
The event-based PI control in \cite{aaarzen1999simple} launched the new and still flourishing field of event-based control; see \cite{aranda2020event} for a recent survey. While the theoretical benefits of event-based over sampled data control are now clearly established, most of the existing literature has concentrated on emulating the behavior of classical digital control systems with an event-based architecture. In contrast, the design of control systems that interconnect rhythmic systems through sensing and actuating events  is still in its infancy \cite{sepulchre2022spiking,fernandez2023neuromorphic}.

The goal of this letter is to  explore the potential of neuromorphic
control with the  simple mechanical model of a pendulum. We regard the pendulum as a rhythmic system and we regard the control problem as the design of another rhythmic system able to orchestrate the behavior of the pendulum. The rhythmic controller is designed in such a way that the desired behavior is achieved by a form of synchrony between the two rhythmic systems. 

The design of a rhythmic event-based controller starts with the
design of an automaton: the desired behavior of the controlled
system is represented by a temporal sequence of sensor and
actuator events. The automaton controller is a rhythmic system
capable of generating the desired sequence of events. The
control objective is formulated as a synchronization problem
between the plant and the automaton. The second step is to
design a regulator that uses output feedback and that tunes continuous parameters to ensure
robustness and adaptability of the controller.

We show that the bio-inspired architecture of neuromorphic controllers such as presented in \cite{ribar2021neuromorphic} 
provides a flexible methodological framework for such a two step design. The neural architecture of the controller consists of simple motifs. The topology of the network defines the automaton, whereas the regulatory properties are achieved by output feedback, and by neuromodulation of the parameters using the classical framework of adaptive control \cite{schmetterling2022adaptive,burghi2023adaptive}. 
Despite being rhythmic and non-linear, the closed-loop system is amenable to rigorous analysis using existing theory. We provide no theoretical development in this letter but point to the relevant references in each section.%\footnote{\color{green} This is stated as a fact; we need to provide a reference.}
%This letter contains no new theory, but application of existing theory to a classical control problem.

The idea of entraining a mechanical system with a neuro-inspired controller has a rich and long history; it underlines much of the literature connecting the central pattern generators of neuroscience and the rhythmic controllers of legged robotics \cite{ijspeert2008central}. Nonlinear oscillators have been sucessfully designed to orchestrate rhythmic behaviors  \cite{ijspeert2007swimming,ijspeert2005simulation}, % ,buchli2006finding}, 
and neuromorphic implementations have received recent attention \cite{angelidis2021spiking}. %Calibration explored in \cite{buchli2006finding}.
A limitation of those approaches is the lack of a systematic modelling framework, which has motivated the design of linear controllers \cite{ryu2021optimality}.
Neuro-inspired controllers have been implemented experimentally, see e.g. \cite{lewis2005cpg,simoni2006two}. But the tuning of those architectures has proven difficult.
Event-based control has been explored for the linearised pendulum \cite{kuo2002relative}, with an approach that utilises classical linear methods. The design of a controller by means of an adaptively tuned rhythmic automaton and an output feedback phase control seems novel.

The letter is organised as follows. The automaton of the plant is introduced in Section \ref{behaviours}. Section \ref{architecture} presents the event-based control architecture; its implementation follows in Section \ref{implementation}. In Section \ref{entrainment}, we control the pendulum in open-loop. Sections \ref{phase} and \ref{adaptive} introduce output and adaptive feedback respectively. Section \ref{conclusion} discusses the potential of the present approach, with the intention of motivating further research from the community.

\section{The automaton of a pendulum} \label{behaviours}

We consider the non-dimensionalised dynamical model of a pendulum 

\begin{equation} \label{eq:dimensionless}
\ddot{q} + \alpha \dot{q} + \sin(q) = I
\end{equation}
where $q$ is the pendulum's angle from the resting position, $\alpha$ is dimensionless damping and $I$ is dimensionless torque.

The rich dynamical behaviour of this seemingly simple system is part of any textbook of nonlinear dynamics see e.g. the excellent treatment in \cite{strogatz1996nonlinear}. With a constant torque, the qualitative behaviour can be comprehensively studied through phase-portrait analysis. 
The key properties of the pendulum dynamics can be summarized as a function of the two parameters $\alpha$ and $I$.

The behavior of the pendulum is simple in the so-called overdamped regime ($\alpha >1$). In this regime, the only attractor of the system is a stable equilibrium, for $I <1 $, or a limit cycle, for $I >1$. 
%In the equilibrium regime, there is a monotone relationship between the equilibrium angle and the amplitude of the torque. In the oscillatory regime, there is a monotone relation between the frequency of oscillations and the amplitude of the torque.
The behavior is more complex in the underdamped regime ($\alpha < 1$). In this regime, the stable equilibrium can coexist with a stable limit cycle, depending on the initial energy of the system. This bistability is the source of complex behaviors, including co-existence between small and large-amplitude
oscillations and sensitivity to initial conditions under non-constant
forcing. Small variations of the applied torque suffice to switch the
behavior between “small” and “large” oscillatory behavior.
%The unpredictability of the long term behavior of the chaotic pendulum amounts to the ultrasensitivity of the switching behavior in the vicinity of the saddle equilibrium point. The interested reader will find a detailed treatment of those dynamical phenomena in \cite{strogatz1996nonlinear}.

In the event-based framework of the present paper, the torque is not constant but  a sequence of impulses of short duration. 
%The analysis of the dynamical behavior under constant inputs is nevertheless a good predictor of the possible behaviors. 
As in the discussion above, we can distinguish between two types of stationary behaviors: \textit{small oscillations} (less than a full swing), that replace the stable equilibrium obtained with a constant torque, and \textit{large oscillations} (sequences of complete rotations), as in the regime of constant torque with $I>1$. This discrete distinction leads to the description of the pendulum as   a two state automaton, with a ``low energy state" corresponding to small oscillations and a ``high energy state" corresponding to large oscillations. Both states can be entrained by a periodic sequence of impulses. Event-based control of the pendulum can be thought as orchestrating the trajectory of that automaton (that is, the switches between the high and low states), as well as tuning each of the states: regulating the frequency and amplitude of the oscillation in the low state, and regulating the frequency of oscillations in the high state. 

\section{Event-based Control Architecture} \label{architecture}

\begin{figure}
\centering
\includegraphics[width=3.5in]{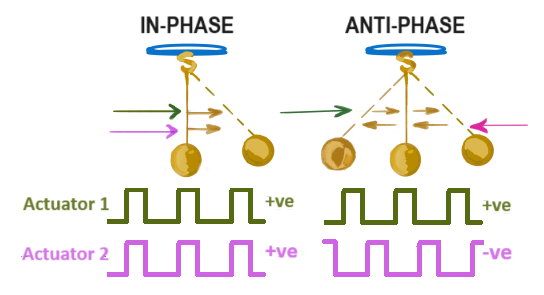}
\caption{The two distinct configurations of event actuation. \textit{Left:} in-phase identical actuating events  (IN-PHASE). \textit{Right:} anti-phase actuating events of opposite sign (ANTI-PHASE). Sketch courtesy of @artjoy2015.}
\label{fig:swing}
\end{figure}

To conceive the rhythmic automaton of the controller,  imagine two children  pushing a swing. By pushing harder or more frequently, they can increase the swing's amplitude or frequency respectively. The children can coordinate in two different ways: they can stand on the same side of the swing and push it in synchrony, or on opposite sides and push it in anti-synchrony. 
We replicate a similar architecture with two independent motors forcing the pendulum. Each actuator  periodically `kicks' the pendulum with a pulse of variable duration. The duration and frequency of the pulses control the amplitude and frequency of the pendulum. Like the children standing on the same side, the actuators can produce a torque of the same sign and act in phase. Or, they can produce torque of the opposite sign and act in anti-phase, `kicking' the pendulum back and forth. We call these two options the ``in-phase configuration" (IN-PHASE) and the ``anti-phase configuration" (ANTI-PHASE) respectively, and we illustrate them in Fig \ref{fig:swing}.

\begin{figure}
\centering
\includegraphics[width=3.5in,trim=0cm 0 0 -3cm]{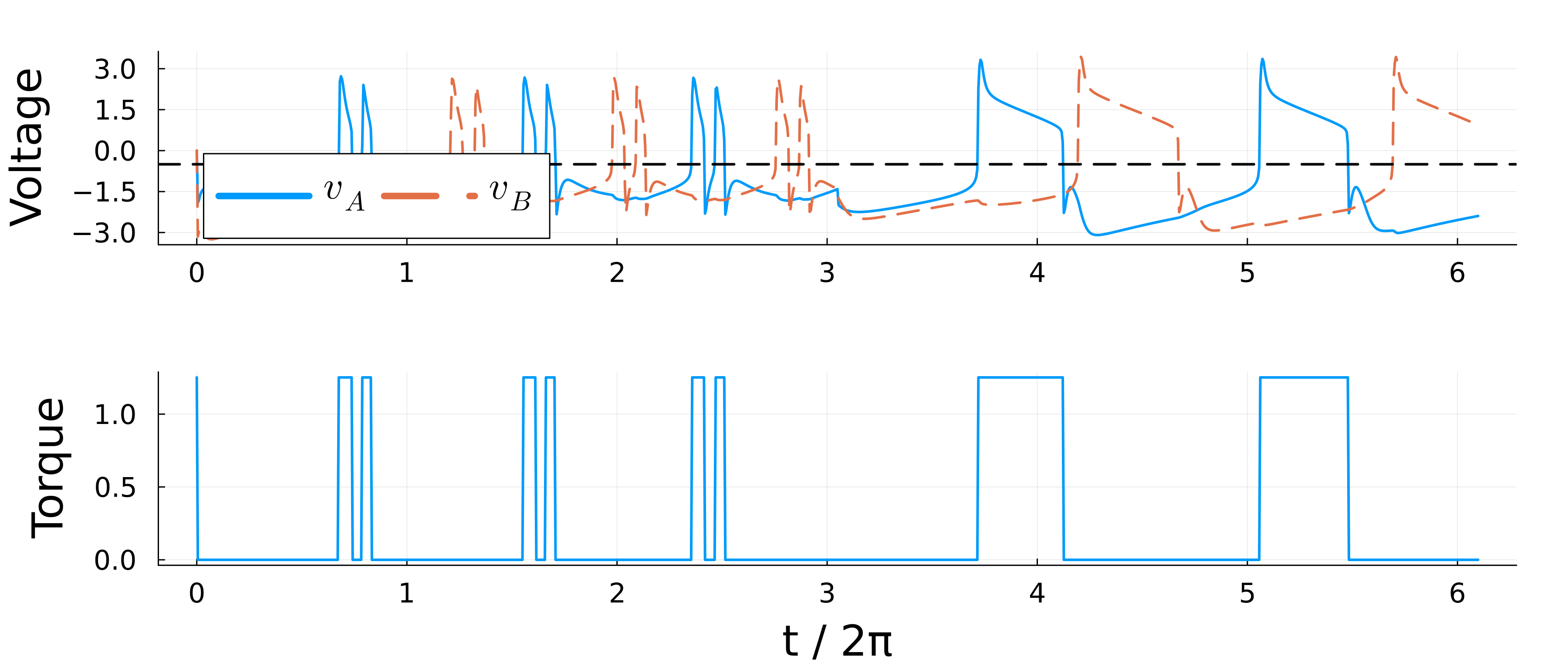}
\caption{The anti-phase rhythm of a Half-Centre Oscillator (HCO) circuit. The size of the bursts is modulated midway. \textit{Top:} the voltages of the two neurons. The dashed horizontal line denotes the voltage threshold, above which the motor is active. \textit{Bottom:} the corresponding actuating events.}
\label{fig:single_HCO}
\end{figure}

In neuromorphic circuits, such rhythms are generated endogenously by simple neuronal motifs, inspired from the central pattern generators of neuroscience  \cite{bucher2015central}. The simplest rhythmic motif is the ``half-centre oscillator" (HCO) \cite{marder2022new}. An HCO comprises two `neurons' (arbitrarily labelled as 'A' and 'B') mutually connected by inhibitory `synapses'. In this letter, we use the HCO model of   \cite{ribar2021neuromorphic}  which was developed for the ease of its tunability. 
%Muscles are often activated by bursting motor neurons \cite{marder2001central}.
Our controller includes one HCO per motor. The HCO drives the motor as follows: when neuron A is spiking, specifically when its voltage is above some threshold, the motor produces a constant torque. When the neuron is below this threshold, the motor is at rest. %\footnote{\color{green}I would remove this sentence. Obtaining an exact value of the torque from a motor requires a feedback loop and often a torque sensor. A constant current output is easier to realize with fast feedback, still the mapping to torque would be nonlinear. I think it is fine to describe the mapping betwen voltage and torque in words, without entering into implementation details.} 
Fig. \ref{fig:single_HCO} shows the behavior of an HCO and the corresponding motor output, for two burst sizes. The burst size controls the actuator event. % The burst size dictates the duration of the torque events; it is governed by the number and duration of the spikes within each burst. % Note that by varying the frequency and size of the voltage bursts, we vary the corresponding properties of the torque wave.

%\begin{figure}
%\centering
%\includegraphics[width=0.2in]{figs/HCO.png}
%\caption{An HCO is a network comprising two neurons (denoted by circles) and two synapses (denoted by arrows).}
%\label{fig:HCO_diag}
%\end{figure}

The coordination between the two motors relies on synaptic interactions between the two HCOs. Excitatory synapses between the pair of A neurons, and also between the pair of B neurons, ensure that the two HCOs are in phase. Inhibitory synapses between the same neurons ensure that the two HCOs are in anti-phase.

%\footnote{\color{red}TODO. (Also I had forgotten to change this sentence to reflect the new figure, sorry.) \color{green} Fig. \ref{fig:network} is confusing. We talk of event-based architecture but we have a figure with Adaptive controller and Phase controller. We talk about neurons A and B, excitatory and inhibitory synapses, but we do not provide any illustration. The neural circuit remains a black box. Perhaps can we 'color' it by making it bigger and adding inside a representation of the neural circuit? If we want to keep the Adaptive and Phase controller boxes of the figure, we should at least introduce them in the text.}. To switch motor configurations we switch the synaptic coupling, and also the sign of one of the motors. Fig. \ref{fig:network-sim} shows one such network switch. The four-neuron circuit determines the automaton of the controller.

\begin{figure}
\centering
\includegraphics[width=3.5in,trim=0cm 0 0 -5cm]{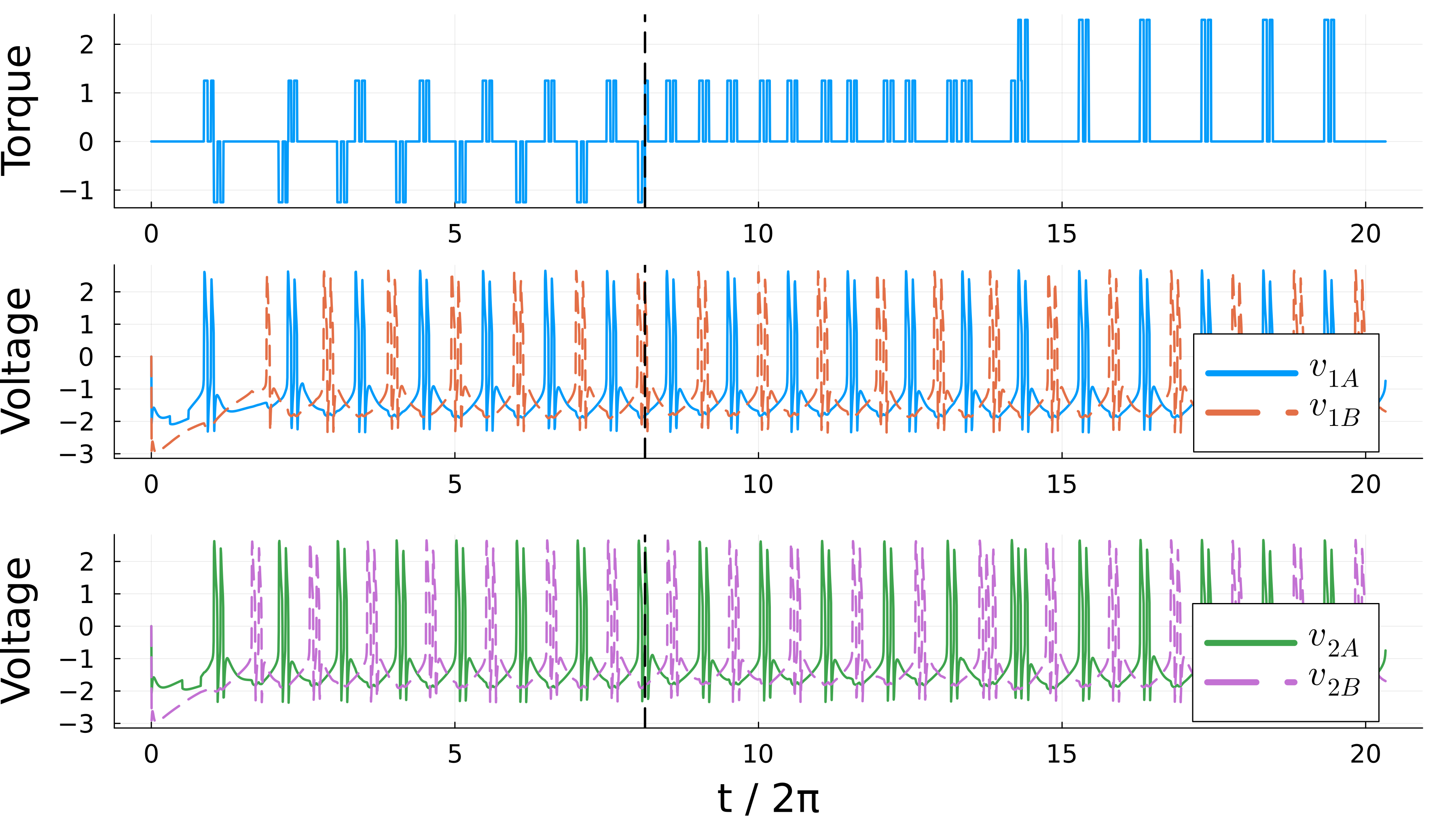}
\caption{Simulation of the HCOs and actuator events. \textit{Top:} the total torque $\torque$. \textit{Middle:} the voltages of HCO 1. \textit{Bottom:} the voltages of HCO 2. At $t = 8$ s (indicated by the dashed black line), the system's configuration is switched from ANTI-PHASE and, after a transient period, it settles on IN-PHASE.}
\label{fig:network-sim}
\end{figure}

\begin{figure}
\centering
\includegraphics[width=3.5in]{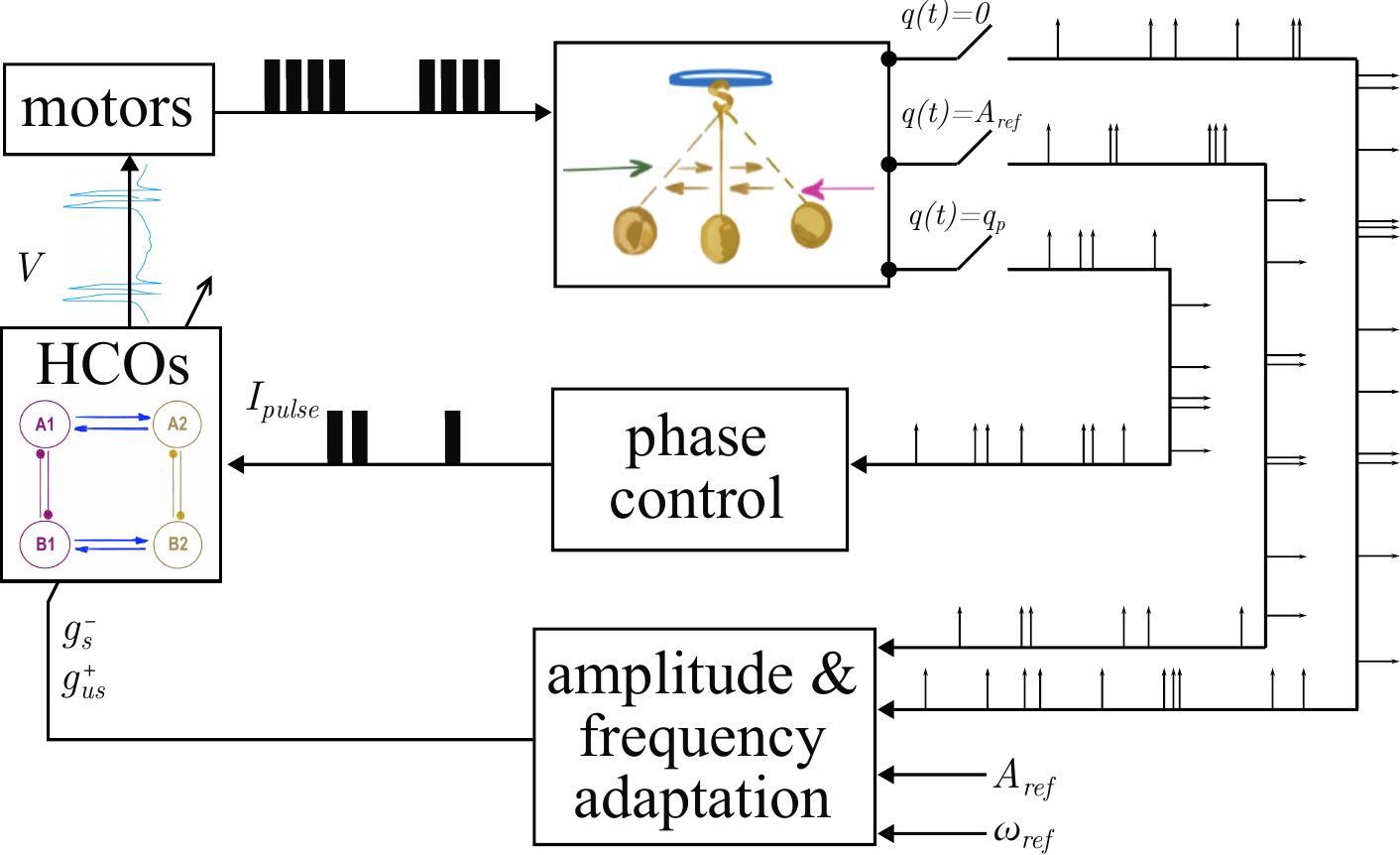} % [width=0.475\textwidth]
%\caption{Diagram of the neuronal network. Neurons are denoted by circles, inhibitory synapses by `disc arrows' and excitatory synapses by `pointed arrows'. The four synapses between the HCOs are either excitatory or inhibitory, for IN-PHASE and ANTI-PHASE respectively.}
\caption{Block diagram of the complete architecture, including the event-based feedback loops introduced in Sections \ref{phase} and \ref{adaptive}. Small arrows over signal transmission lines indicate event-based communication as described in Section~\ref{architecture}. The HCO block architecture is described in Sections~\ref{architecture} and~\ref{implementation}.
%Quantities in round brackets are events. The neuronal network is also shown. The circles denote neurons. The `disc arrows` denote inhibitory synapses. The `pointed arrows` denote synapses that are either inhibitory or excitatory (for ANTI-PHASE and IN-PHASE respectively).
}
\label{fig:network}
\end{figure}

The frequency and amplitude of the pendulum oscillations are estimated from event-based sensors. In this letter, we use three photodetectors that detect specific angle crossings: one for the output feedback phase controller ($q=q_p$), and two for the adaptive regulation of frequency and amplitude ($q=0$ and $q=A_{\rm ref}$). 

Fig. \ref{fig:network} shows the block diagram of the complete architecture.% and the neuronal network.

\section{Neuromorphic implementation} \label{implementation}

The  circuit realization of our rhythmic controller follows the mixed-feedback motif of spiking control systems \cite{sepulchre2022spiking}. Such systems are minimum phase, relative degree one, and linear in the control parameters.  Those properties make the controller ideally suited for output feedback control and for adaptive regulation of the control parameters. The interested reader is referred to   \cite{burghi2023adaptive,schmetterling2022adaptive}.

Following the approach in \cite{ribar2019neuromodulation}, %which provides an easy relationship between the neuron parameters and the tuning of the bursting behavior.  
%This neuron is an RC circuit in parallel with a bank of current sources, with each current source on a separate parallel branch.
each neuron is an RC circuit in parallel with a bank of voltage-controlled current sources.
Every current source provides either positive or negative conductance, at a particular timescale. A positive conductance is a source of negative feedback, and vice versa. The gains of the current sources are the control parameters.
Each neuron $i$ has three state variables: a voltage $v_i$ (the output) and two low-pass-filtered voltages $v_{s,i}$ and $v_{us,i}$ (respectively `slow' and `ultra-slow' voltages). The dynamics are:
\begin{align*}
\tau_f \dot v_i &= -v_i
+ \afn \tanh(v_i) - \asp \tanh(v_{s,i}) \\ &+ \asn \tanh(v_{s,i} + 0.9) - \ausp \tanh(v_{us,i} + 0.9) \\ &+ I_{\rm{syn},ij} + \iappi \\
\tau_s \dot{v}_{s,i} &= v_i - v_{s,i} \\
\tau_{us} \dot{v}_{us,i} &= v_i - v_{us,i}
\end{align*}
% for $i, j \in [1,2]$ and $i \neq j$. \textbf{Todo: correct the index pairs for the synapses.}
for $i \in \{A1,B1,A2,B2\}$. The ultra-slow timescale $\tau_{us}$ is chosen to (roughly) align with the natural frequency of the pendulum. This nominal choice enables a low-power controller to achieve a wide range of amplitudes by resonance. The other timescales are chosen such that they are all sufficiently separated \cite{ribar2019neuromodulation}. 

Following the design procedure in \cite{ribar2021neuromorphic}, a neuron circuit can be manually tuned using its three current-voltage curves (one at each timescale); these curves also demonstrate that the behaviour is quite robust to parameter uncertainty \cite{ribar2019neuromodulation}.

For the synapses we follow  \cite{ribar2020synthesis}: the synaptic current from neuron $j$ to neuron $i$ obeys $$I_{\rm{syn},ij} = g_{\rm{syn},ij} / (1 + \exp(-2(v_{s,j} + 1)))$$ 
where $g_{\rm{syn},ij}$ is positive for an excitatory synapse, and negative for an inhibitory one. 

For simplicity, we use identical parameters for the four neurons as well as for each type of synapse. The controller behavior is however quite robust to heterogeneity in the parameters. 
The neuron's input is the applied current $\iappi$. We set $\iappi$ to a nominal value $\iappnominal = -1$. The controller does not oscillate spontaneously, but a brief pulse of current in any of the neurons is sufficient to turn the rhythm `on'.  A full list of the parameter values used in each figure is available with the attached code.\footnote{https://github.com/RJZS/neuromorphic-pendulum-control} All quantities in this letter are non-dimensional; see \cite{ribar2019neuromodulation} for the circuit implementation.
%Fig. \ref{fig:hco_rhythms} provides some illustrative examples of the HCO oscillating in anti-phase.

The only tuning parameters in this letter are the gain of the slow negative conductance  $\asn$  and the gain of the ultraslow positive conductance $\ausp$. Those two parameters respectively modulate the burst size and the burst frequency.  All the other parameters are fixed throughout the paper.

\section{Control by entrainment} \label{entrainment}	

%\begin{figure}
%	\centering
%	\includegraphics[width=3.5in]{figs/0_hco_rhythms.png}
%	\caption{Example oscillations of the HCO, showing the output voltages $v_1$ (blue) and $v_2$ (red). From top row to bottom row: by increasing $\asn$, we can increase the duty cycle. We must also adjust $\ausp$ to preserve the frequency. The wide spikes in the bottom row are not physiological, but they are useful for pendulum control.}
%	\label{fig:hco_rhythms}
%\end{figure}

Feedforward control of a rhythm by another rhythm is commonly referred to as  {\it entrainment}. The role of entrainment in biology can be regarded as analog to the role of resonance in mechanics.  It is widely observed, and it is  robust to uncertainty in that it does not require an exact prior match between the frequencies of the input rhythm and the entrained rhythm \cite{lakatos2019new}. 
Any stable limit cycle is locally entrainable by an entraining signal of small amplitude \cite{wieland2010phase}.  Global entrainment has also been studied for contractive systems \cite{Sontag2010,Russo2010}.% The input frequency does not need to match the natural frequency of the entrained system.

As a first step, we consider entrainment at a fixed frequency (chosen to be close to the natural frequency of the pendulum) by a feedforward controller. 
%For small oscillations, we consider the anti-phase configuration of the controller. 

\begin{figure}
\centering
\includegraphics[width=3.5in,trim=0cm 0 0 -5cm]{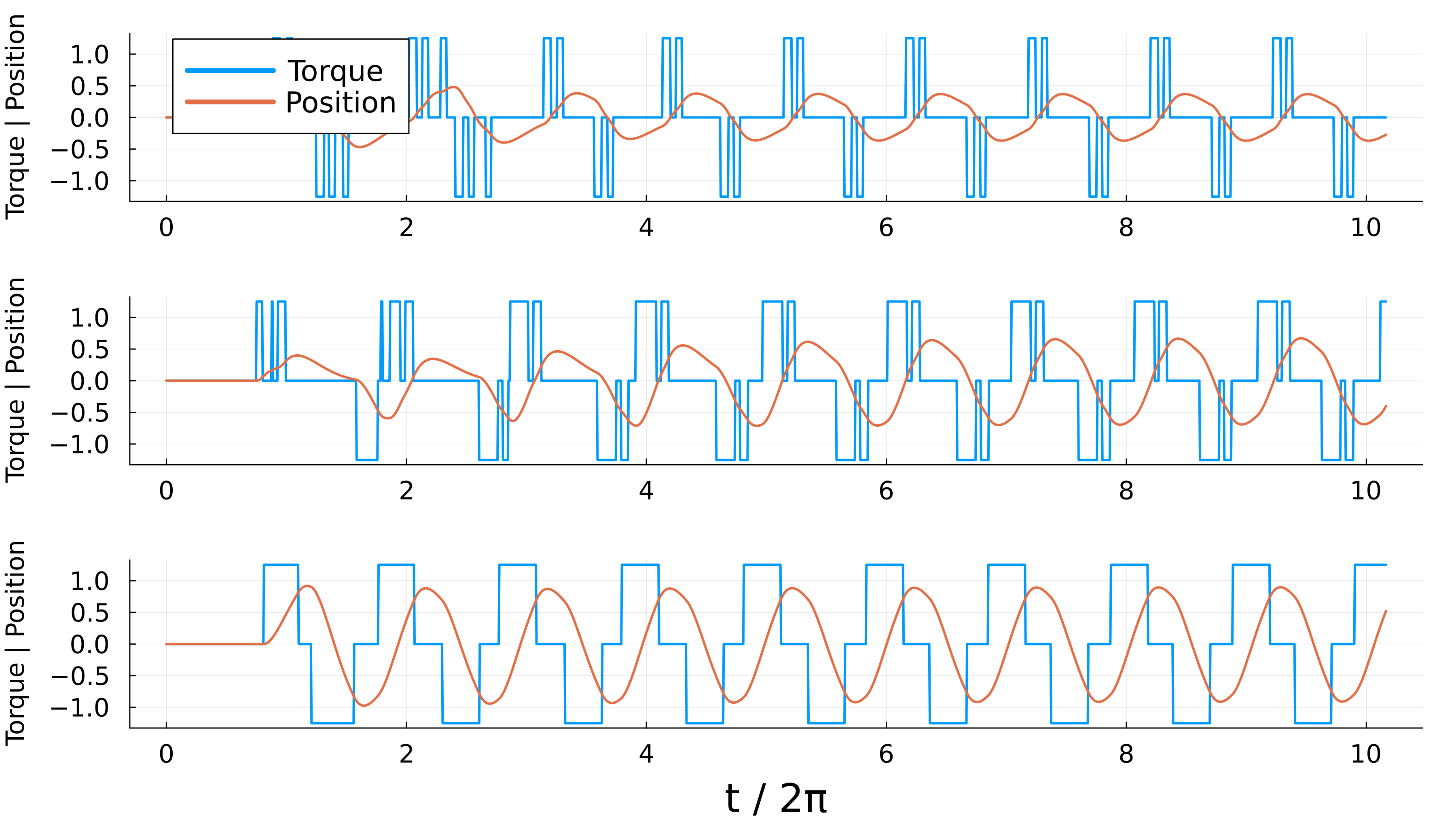}
\caption{Small oscillations in the overdamped regime (damping $\alpha = 1.4$). As the burst size of the neural oscillation increases (from top row to bottom row), so does the amplitude of the pendulum's oscillation.}% The neural parameters were determined in advance and are held constant throughout each simulation.}
\label{fig:overdamped_regime}
\end{figure}

\begin{figure}
\centering
\includegraphics[width=3.5in,trim=0cm 0 0 -0.5cm]{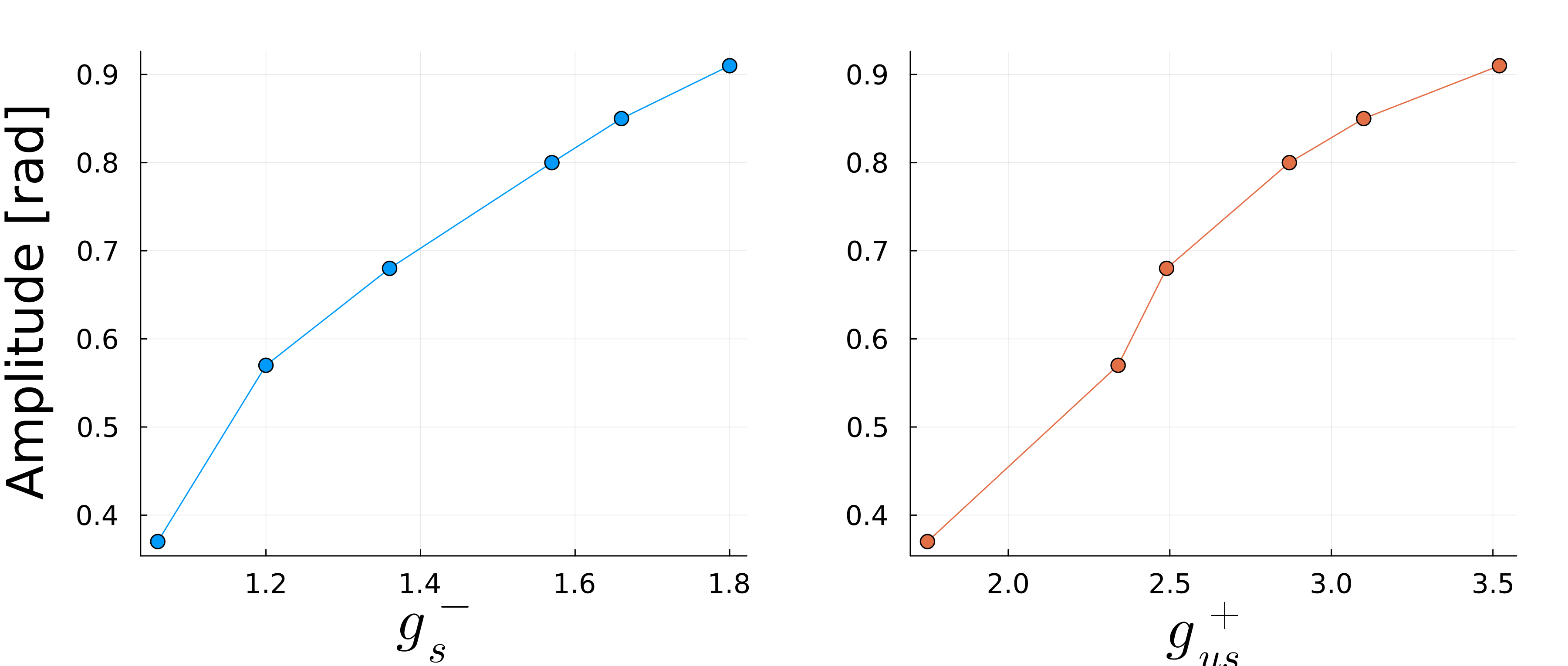}
\caption{Entrainment of the pendulum at a fixed frequency in the overdamped regime ($\alpha = 1.4$). The amplitude of small oscillations increases monotonically with $\asn$ and $\ausp$. }
\label{fig:overdamped_monotone}
\end{figure}

Fig. \ref{fig:overdamped_regime} shows entrainment of the overdamped pendulum at the fixed frequency, for different choices of burst size and using the anti-phase configuration of the controller. The amplitude increases with burst size, % Energy too.
which is regulated by $\asn$. This gain also affects the frequency of the HCO, a variation compensated by the parameter $\ausp$. Fig. \ref{fig:overdamped_monotone} illustrates the monotonic relationship between the amplitude of the entrained pendulum and the neural gains $\asn$ and $\ausp$ at the fixed frequency. % We consider the behaviour in this regime as a PWM-like 'rate code' paralleling that of the neuron servo \cite{deweerth1991simple}.

% In the overdamped regime, Figure illustrates the  monotone relationship between the energy injected into the system and the amplitude of oscillations.
% This is consistent with the dynamical analysis in Section \ref{behaviours} 

\begin{figure}
\centering
\includegraphics[width=3.5in]{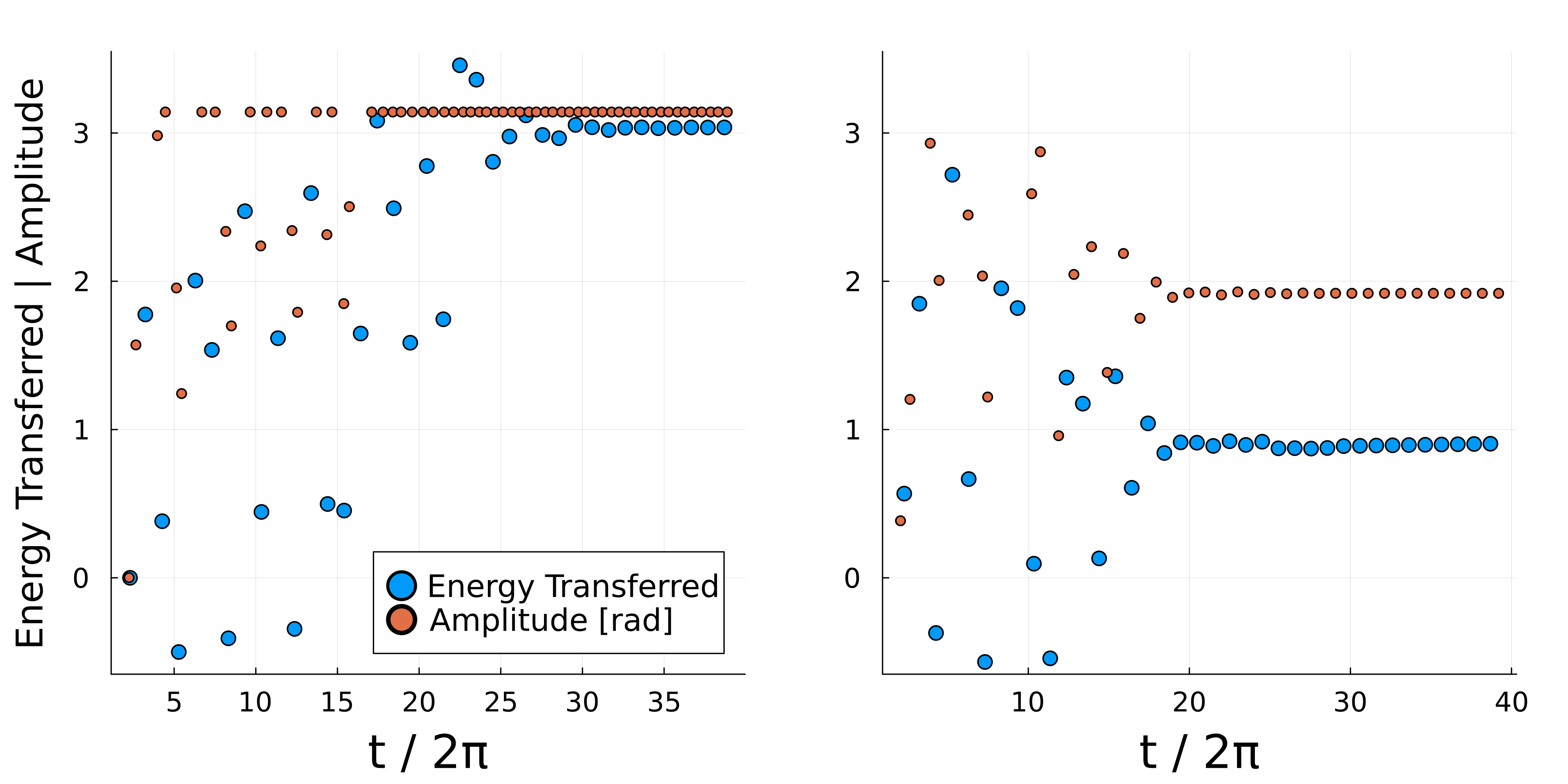}
\caption{Entrainment with an IN-PHASE neural oscillation in the underdamped regime (in this example, $\alpha = 0.14$), showing energy transferred by the motors each burst. The amplitude of each oscillation is also shown (with an amplitude of $\pi$ indicating a complete swing). Depending on the initial conditions, the steady-state oscillations are either large or small (left and right plots respectively).}
\label{fig:bistability}
\end{figure}

% Consistent with this analysis, 
The entrainment behavior becomes more complex in the underdamped regime: IN-PHASE entrainment leads to one of two possible stable states (small or large oscillations), depending on the initial conditions. Fig. \ref{fig:bistability} shows the energy transferred by the motors during each neural period, $E_i = \int_{t_i}^{t_i+T} \dot{q} I dt$, where each time $t_i$ is at the beginning of a burst event, and $T$ is the time until the next such event. The figure shows that the pendulum can converge to either the high or the low energy state. The high-energy state in Fig. \ref{fig:bistability}  is a 2:1 oscillation, meaning  two complete swings of the pendulum for every actuator event.

%\begin{figure}
%	\centering
%	\includegraphics[width=3.5in]{figs/5\_switching\_bistability.png}
%	\caption{A switch from ANTI-PHASE to IN-PHASE. Starting in small oscillations, the pendulum after the switch settles into either small (top row) or large (bottom row) oscillations, depending on the switching time (denoted by the dashed vertical line).}
%	\label{fig:bistability_example} 
%\end{figure}

% Fig. \ref{fig:overdamped_regime} shows entrainment at 1 Hz with the anti-phase configuration, for different values of $\asn$.

%\begin{figure}
%	\centering
%	\includegraphics[width=3.5in]{figs/1\_overdamped.png}
%	\caption{Small oscillations in the overdamped regime. Torque $\tau$ and angle $q$ are in Nm and rad respectively. As the kick size of the neural oscillation increases (from top row to bottom row), so does the amplitude of the pendulum's oscillation. The neural parameters were learned in advance and are held constant throughout each simulation.}
%	\label{fig:overdamped_regime}
%\end{figure}

\section{Phase Control} \label{phase}

In contrast to small oscillations, the entrainment of large oscillations in the underdamped regime is {\it fragile}. A slight change in the parameters can lead to a non-periodic behaviour, which again is consistent with the chaotic attractor that the pendulum can exhibit with a small sinusoidal input superposed to a constant torque input. In this section, we use  feedback control to significantly enlarge the basin of attraction of the large oscillations.  

\begin{figure}
	\centering
	\includegraphics[width=3.5in]{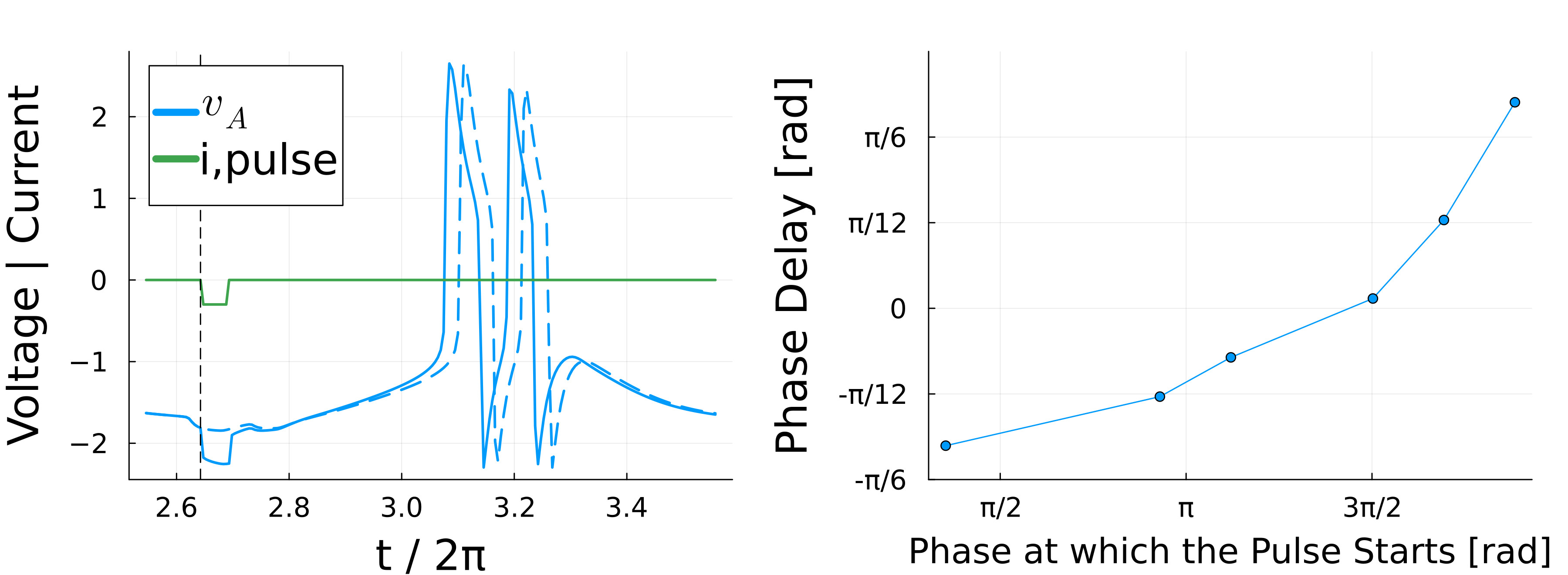}
	\caption{\textit{Left:} A small inhibitory pulse (green) causes a phase advance or delay of the subsequent burst. The blue curve shows $v_A$ for an isolated HCO. The blue dashed curve shows the behaviour of $v_A$ in the absence of the pulse. \textit{Right:} Phase response curve of an HCO. The curve indicates the phase shift resulting from a fixed pulse ($w = 0.05$ and $P = 0.3$) as a  function of the timing of the pulse along the limit cycle.}
	\label{fig:phasebehaviour}
\end{figure}

\begin{figure}
\centering
\includegraphics[width=3.5in]{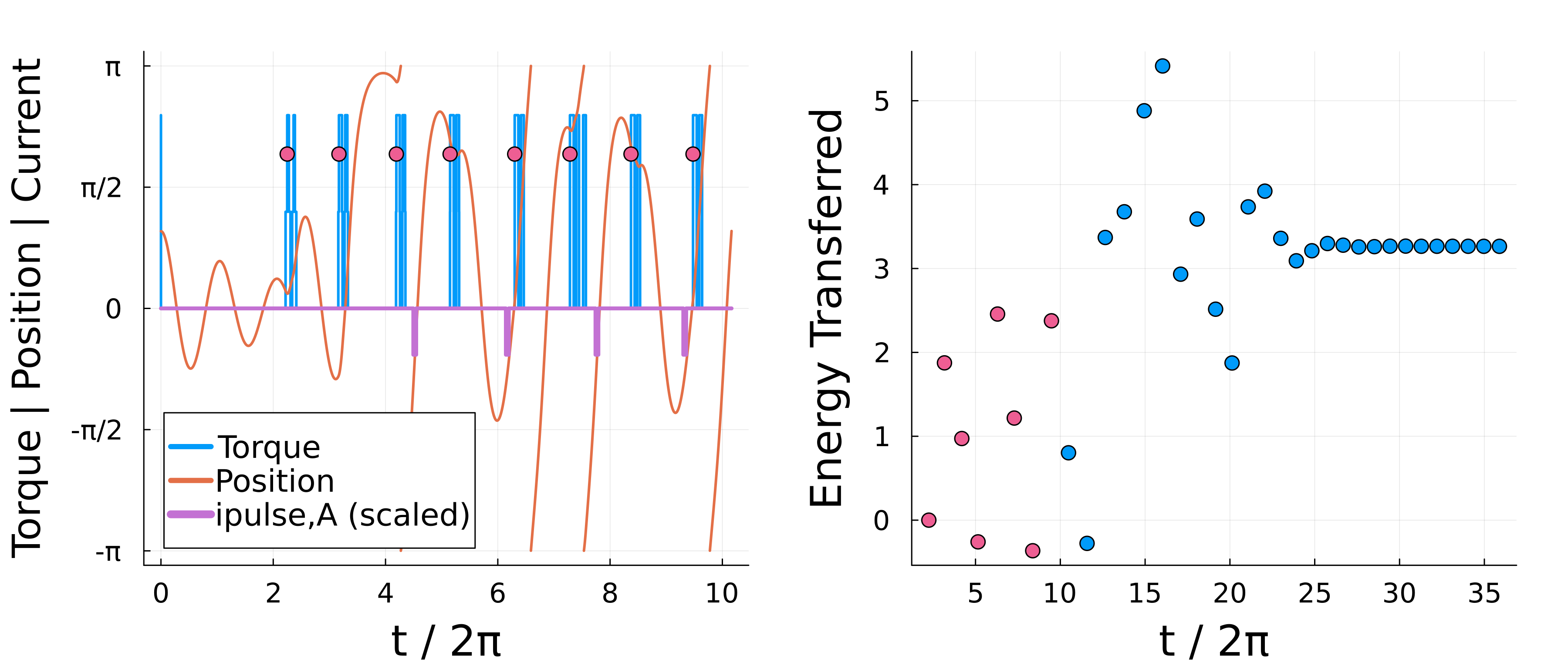}
\caption{The simulation of Fig. \ref{fig:bistability} (right plot) with the addition of phase control, forcing convergence to the high- rather than the low-energy state. \textit{Left:} the initial response. The current pulses to the A neurons are shown in purple, and are scaled for clarity. Each pink marker indicates the start of an actuation event. \textit{Right:} the energy transferred per event, $E_i$, showing convergence to a steady-state oscillation. The pink markers match those of the left plot.}% $w = 0.05$ s and $P = 0.3$.}
\label{fig:phasecontrol-tuning}
\end{figure}

To design an event-based output feedback control, we use the concept of phase control. Phase control consists of advancing or delaying the next actuator event via small negative pulses added to the nominal (constant) input current of each neuron.   The phenomenon is illustrated in Fig. \ref{fig:phasebehaviour} (left). The  delay or advance of the next actuator event  is a function of the timing of the pulse, a relationship known as the `phase response curve' (PRC) \cite{sacre2014sensitivity}. It is known that a monotone phase response curve allows control of the phase of an oscillator by a simple proportional controller; see \cite{sacre2014sensitivity,efimov2009controlling}. The PRC method is based on linearized analysis, which  assumes  {\it small} perturbations. We use perturbations of fixed amplitude (P) and fixed duration (w). Fig. \ref{fig:phasebehaviour} (right) illustrates the strong monotonicity of the  PRC of the HCO oscillator, making the simple proportional phase control effective. % for a nominal oscillation at the fixed frequency.  
The phase controller %{\color{green}proportional feedback controller}\footnote{\color{green} The feedback? The phase controller?} 
applies a current pulse at each sensory event $q(t) = \qsensory$.  The parameter $\qsensory$ (selected to $\qsensory = -1$ rad in the simulation of Figure \ref{fig:phasecontrol-tuning}) can be regarded as the selection of a proportional gain. It 
does not require a precise tuning.  In the IN-PHASE configuration, the feedback pulses are at each sensory event injected alternately into the two A neurons and the two B neurons.

Fig. \ref{fig:phasecontrol-tuning} illustrates the stabilizing property of a proportional phase control. The initial conditions are those of Fig. \ref{fig:bistability} (right plot), but the phase control forces convergence to the high- rather than low-energy state. 
The steady-state frequency is slightly different to that obtained with feedforward entrainment, and varies with $\qsensory$.
%The figure also shows that, by varying $w$, we can tune the frequency. %Fig. \ref{fig:phasecontrol} plots the transient region around the time of the switch, to better illustrate how the phase control kicks the pendulum into the high state. 

%$P$ and $w$ must be small for the monotonic PRC to hold (else the pulses will impact the burst size). 
%We should note that in order to prevent braking during the transient, a sensory event is ignored if the pendulum's velocity is negative. 

\section{Neuromorphic Adaptive Control} \label{adaptive}

Online adaptation of the control parameters can be used as well as phase control in order to regulate an oscillation.

%As a first step, the sensory information is only used to {\it adapt} the controller parameters. We illustrate this by regulating the frequency $\omega$ and amplitude $A$ of small oscillations through adaptive control of the neural gains $\ax$. 
Robust adaptive regulation of the control parameters has been previously studied in \cite{schmetterling2022adaptive,burghi2023adaptive,schmetterling2023robust,burghi2022distributed}. This work relies on the property that the proposed neuronal architecture inherently satisfies the standard assumptions of adaptive control: the system is minimum phase, relative degree one, and linear in the parameters. Adaptive control is analogous to the biological concept of neuromodulation \cite{marder2014neuromodulation}.

%Here, we use adaptive control to learn gains that give the desired frequency and amplitude of small oscillations.
%Automated tuning is particularly important due to 'transistor mismatch', a major issue facing neuromorphic engineers \cite{liu2010neuromorphic}. This mismatch is a constant component error that is introduced into transistors at the point of manufacture, which means that nominally identical circuits will in fact have different parameters.

We consider the online adaptation of the two parameters $\ausp$ and $\asn$, which regulate the neural frequency and burst size respectively. We use these to regulate the frequency and amplitude of small oscillations in ANTI-PHASE. The online adaptation is implemented via the integral action of a prediction error computed from the sensory events.

For a reference frequency  $\wref$, we predict that a zero crossing  $q(t)=0$ will occur  every $\pi/\wref$ seconds. A  correction $\pw$ is added to the parameter $\ausp$   at each zero crossing, with a sign determined by the sign of the prediction error and an amplitude proportional to its magnitude. 

Likewise, for  a reference amplitude $A_{ref}$
 for the pendulum's oscillation,  we predict that an event $q = \Aref$ occurs  $\pi/2\wref$ seconds after each $q=0$ event.
If no such event occurs before the next zero crossing, we apply a fixed correction $\pA$ to the parameter $\asn$. Otherwise, the correction is applied at the event $q = \Aref$ with a sign and amplitude determined by the prediction error. 

The simple adaption law described above can be regarded as  a phase integral control. In practice, the parameter variation is confined to a finite range such that bursting is preserved.

%We also introduce the error signals $\ew = \freq - \wref$ and $\eA = \amp - \Aref$. We adapt the two parameters $\ausp$ and $\asn$ which regulate the neural frequency and burst size, respectively.

\begin{figure}
\centering
\includegraphics[width=3.5in]{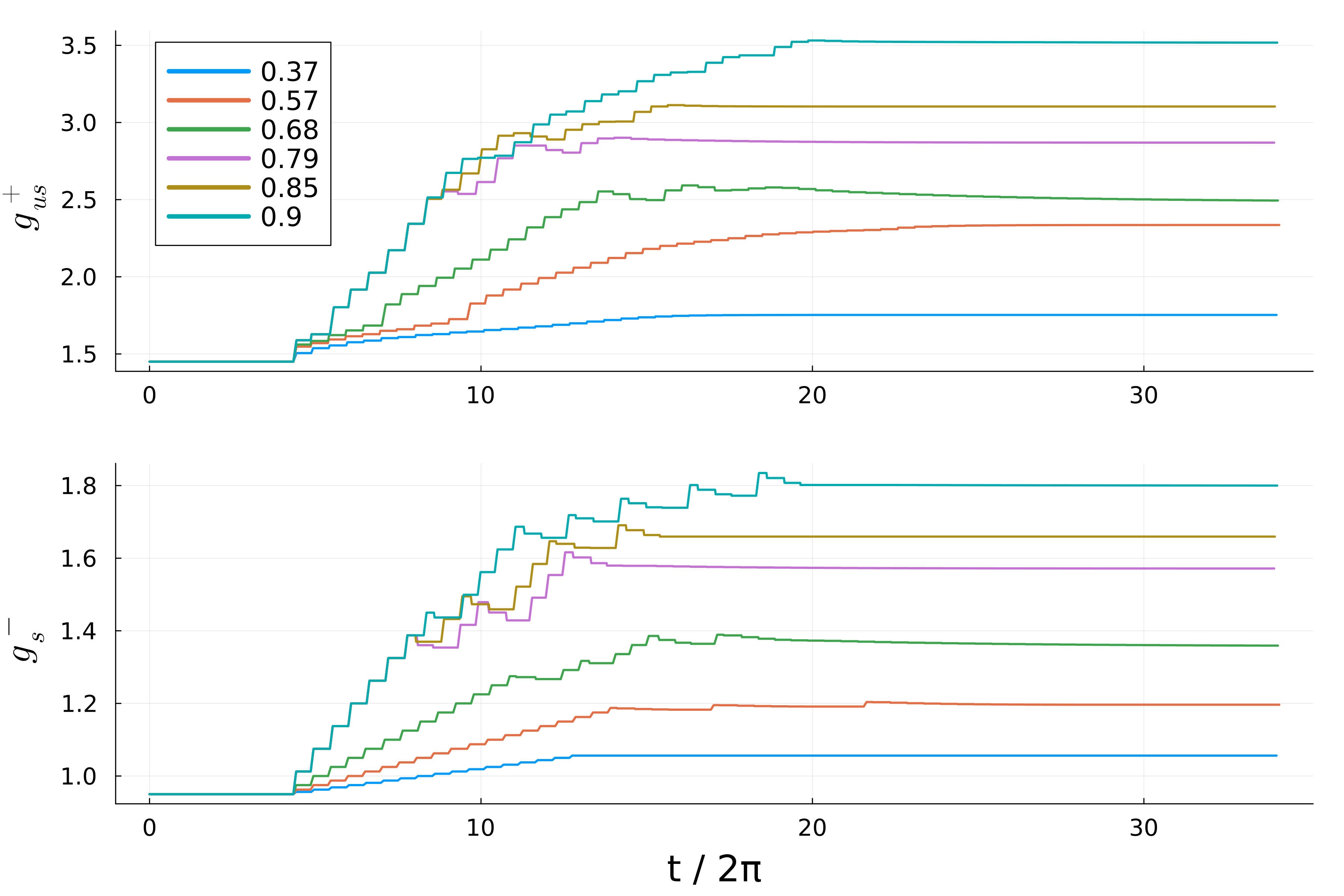}
\caption{The gains tuned by the adaptive controller, when used to obtain the parameters in Fig. \ref{fig:overdamped_monotone}. The legend gives the values of $\Aref$ in radians. In each case, $\wref$ is the fixed frequency chosen in Section \ref{entrainment}.}
\label{fig:tuning}
\end{figure}

Fig. \ref{fig:tuning} illustrates parameter convergence for the reference values illustrated in Figure \ref{fig:overdamped_monotone}.   % The adaptive controller also reduces the sensitivity of the amplitude to perturbations such as a change in mass. 
%(Due to the entrainment, the frequency is already quite robust in open-loop.) 

\section{Discussion} \label{conclusion}

We have presented a neuromorphic, event-based framework for controlling the oscillations of a mechanical pendulum. A simple neuromorphic controller was used to generate actuator events capable of entraining the pendulum oscillation while regulating its energy. The feedforward controller exhibits bistability between low-energy and high-energy oscillations in the underdamped regime. A proportional phase control was designed to stabilise the high energy oscillation only. Finally, an integral phase control was designed to  regulate oscillations at a given frequency and amplitude. While no formal convergence analysis was included in the letter, the design was supported by existing analysis and system properties that enable  regulation by simple feedback controllers (minimum phase and relative degree one).   % We therefore consider the complete control architecture as an event-based PI controller for rhythmic systems.

%The event-based nature of sensing and actuation enables the distinct design of an automaton able to generate the required actuation events and a regulator able to tune the steady-state behaviour. In turn, bio-inspired neural architectures offer a versatile framework to design the automaton through specific network arrangements of  basic motifs and to tune the automaton through either neuromodulation, that is, adaptive regulation of the control parameters, or phase control, that is, modulation of the events' relative timings.

The event-based nature of the designed controller offers a number of potential advantages. Prime and foremost, the energy exchange between the systems is confined to the events. As a consequence, the temporal sparsity of the events is a direct measure of the energy efficiency of the design. High impedance control during the events is an inherent source of robustness to model uncertainty. Temporal sparsity of the events simultaneously ensures low impedance control when averaged over time. In this way, event-based control could potentially combine the benefits of soft actuation with the high impedance requirements of robust control, suggesting a path to overcoming the classical  stiffness-compliance trade-off \cite{pratt1997stiffness}. 
%Furthermore, the event-based control strategy is inherently robust to model and controller uncertainties because of the high impedance but highly localised-in-time nature of the interactions. This pulse-based approach to actuation suggests a path to overcoming the stiffness-compliance trade-off \cite{pratt1997stiffness}. 
The considered  two-motor architecture  also suggests how actuation events can be easily distributed over possibly many actuators, with phase relationships ensured by the architecture of the controller.

%The goal of this letter was to show that the design of the automaton of the controller considerably facilitates the design and  tuning of the feedback control laws. There is certainly room for more advanced design of the adaptation and feedback rules, but the goal was to demonstrate the benefits of entrainment and synchrony for the design of simple and robust feedback control. 

The purpose of the letter was to illustrate the potential of neuromorphic control in a simple mechanical system. We hope that it will stimulate further theoretical and practical research in the control of  robotic or biomechanical rhythms. It would seems natural to explore the potential of neuromorphic control architectures in robotic designs that have already exploited rhythmicity, for instance   \cite{ijspeert2007swimming}. 

\addtolength{\textheight}{-2cm} % Reduced due to compile error - Raph 14/03/22.

%%%%%%%%%%%%%%%%%%%%%%%%%%%%%%%%%%%%%%%%%%%%%%%%%%%%%%%%%%%%%%%%%%%%%%%%%%%%%%%%

%%%%%%%%%%%%%%%%%%%%%%%%%%%%%%%%%%%%%%%%%%%%%%%%%%%%%%%%%%%%%%%%%%%%%%%%%%%%%%%%


%	
% \section*{ACKNOWLEDGMENT}

%%%%%%%%%%%%%%%%%%%%%%%%%%%%%%%%%%%%%%%%%%%%%%%%%%%%%%%%%%%%%%%%%%%%%%%%%%%%%%%%

\bibliographystyle{IEEEtran}
\bibliography{cdc_refs}

\end{document}